\documentclass[final,1p,times]{elsarticle} 

\usepackage{graphicx}
\usepackage{amssymb} 
\usepackage{amsthm} 
\usepackage{lineno}

\journal{Nuclear Physics A} 

\begin{document}

\begin{frontmatter} 

\title{Forward+near-forward azimuthal correlations in p+p and d+Au collisions at $\sqrt{s_{NN}} =
200$~GeV}

\author{Xuan Li (for the STAR\fnref{col1} Collaboration)}
\fntext[col1] {
Now at Physics Department, Temple University, 1801 N. Broad Street · Philadelphia, PA, 19122, USA. 
}
\address{Physics Department, Shandong University, 27 Shanda South Road, Jinan, Shandong 250100, China.}


\begin{abstract} 
Forward particle production and correlation measurements at RHIC can probe low-$x$ gluons. The suppression observed in back-to-back forward $\pi^{0}$+forward $\pi^{0}$ correlations at STAR in central d+Au collisions at STAR is consistent with a prediction of the Color Glass Condensate (CGC) calculation, indicating the gold nucleus probed at such low-$x$ is in a dense gluon state. The forward $\pi^{0}$ + near-forward jet-like cluster azimuthal correlations in 200 GeV p+p and d+Au collisions at STAR are studied, which are sensitive to the intermediate x region between forward+mid-rapidity correlations and forward+forward correlations. Together with the other measurements from STAR, which probe different regions of x, forward+near-forward di-hadron correlations can provide information to understand how sharp is the transition from dilute parton gas to dense gluon state.
\end{abstract} 

\end{frontmatter} 


\section{Introduction}
Gluon saturation is expected at low-$x$ region, where gluon recombination balances gluon splitting. The nuclear gluon distribution is expected to be denser than the nucleon gluon distribution, since the saturation scale $Q_{s}^{2}$ is expected to increase as $A^{1/3}$ \cite{refimpact}. To select a certain low-$x$ region, di-hadron azimuthal correlations triggered on the leading particle in the forward rapidity are studied in d+Au collisions to probe the gluon distribution function of Au nuclei. According to a PYTHIA simulation study, the pseudorapidity of the associated particle is strongly correlated with the soft gluon x involved in the asymmetric partonic scattering \cite{fms_propasal}. With RHIC Run 8 $\sqrt{s_{NN}}$ = 200~GeV p+p and d+Au collision data, the STAR experiment has previously measured forward+mid-rapidity and forward+forward azimuthal correlations. Significant broadening from p+p to d+Au collisions is found in the back-to-back forward+forward correlations. This phenomenon is not observed in the forward+mid-rapidity correlations \cite{ermes_thesis}. Suppression of the forward+forward back-to-back correlations can be described by calculations based on the Color Glass Condensate \cite{for_for, for_cgc}. Forward+near-forward correlations probe the intermediate x region between forward+mid-rapidity and forward-forward correlations \cite{xuanli_thesis}. This measurement would complete the full picture of the transition from a dilute parton gas to a saturated parton density for gold nuclei. 

\section{Experimental setup and event reconstruction}
The STAR experiment at RHIC now has nearly continuous electromagnetic calorimetry spanning pseudorapidity $-1 < \eta < 4$ with full azimuthal angle coverage. It consists of a Barrel ElectroMagnetic Calorimeter (BEMC) with coverage over the interval $-1 < \eta < 1$, an Endcap ElectroMagnetic Calorimeter (EEMC) with coverage over the interval $1.08 < \eta < 2$ and an added Forward Meson Spectrometer (FMS) with coverage over the interval $2.5< \eta < 4$. In RHIC Run 8 d+Au collisions, the deuteron beam heads to the FMS detector. For the forward+near-forward correlations, events require an FMS trigger and a reconstructed $\pi^{0}$ in the FMS. Neutral pions are surrogates for low $p_{t}$ jets and can be used to constrain the initial-state kinematics. Alternatively, associated $\pi^{0}$ can be reconstructed in the EEMC, but the statistics are limited for $\pi^{0}$ reconstructed with EEMC towers \cite{xuan_wwnd}. So, in this work, a jet-cone algorithm with a cone radius $R=\sqrt{\Delta\eta^{2}+\Delta\varphi^{2}}$ of 0.6 is applied to reconstruct jet-like clusters in the EEMC. Jet-like clusters must satisfy a mass cut ($M > 0.2$~GeV/c$^{2}$) for this analysis. The mass cut aims to reduce inclusive $\pi^{0}$ contribution to jet-like clusters.

\begin{figure}[htbp]
\begin{center}
\includegraphics[width=0.48\textwidth]{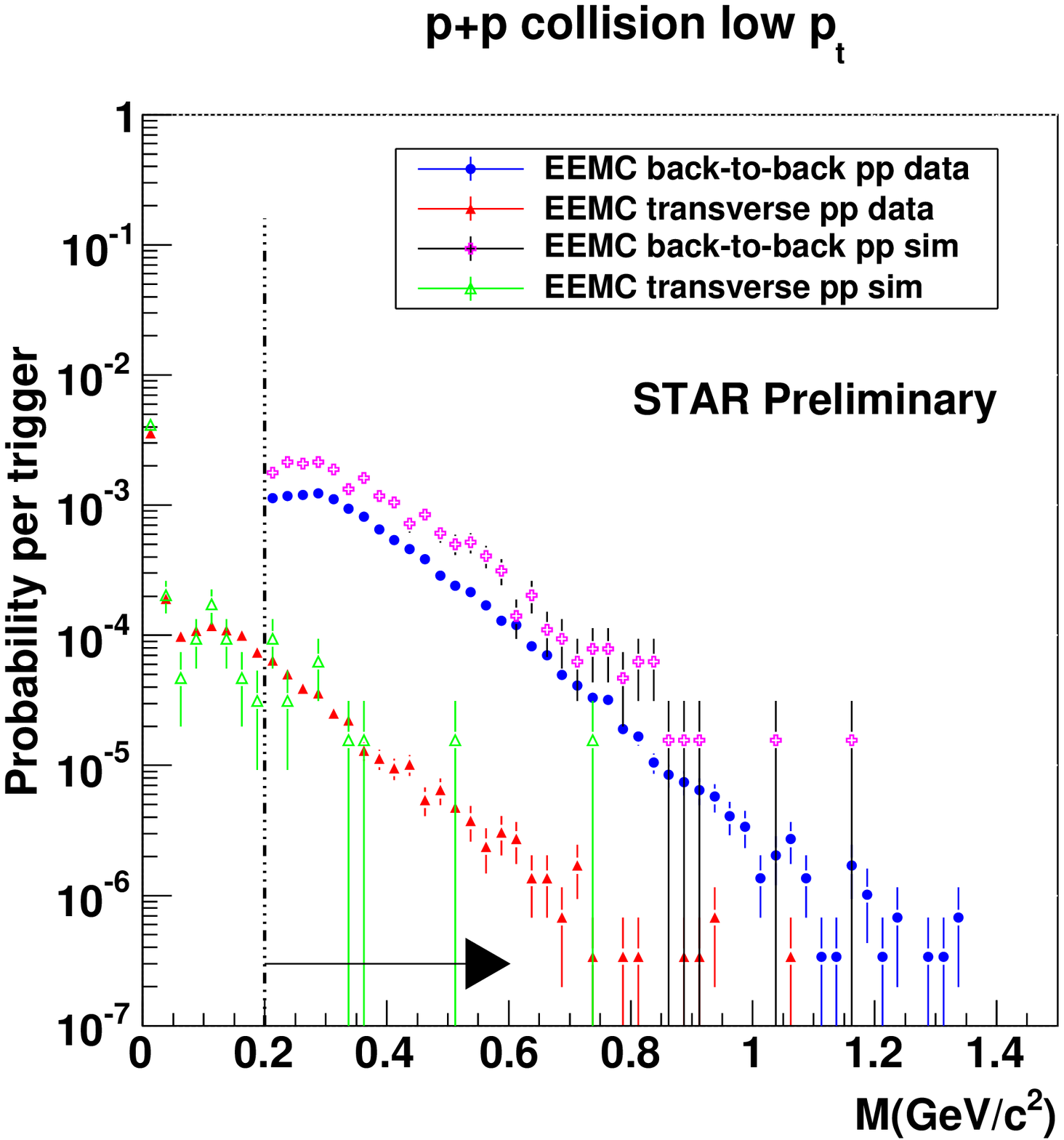}
\includegraphics[width=0.48\textwidth]{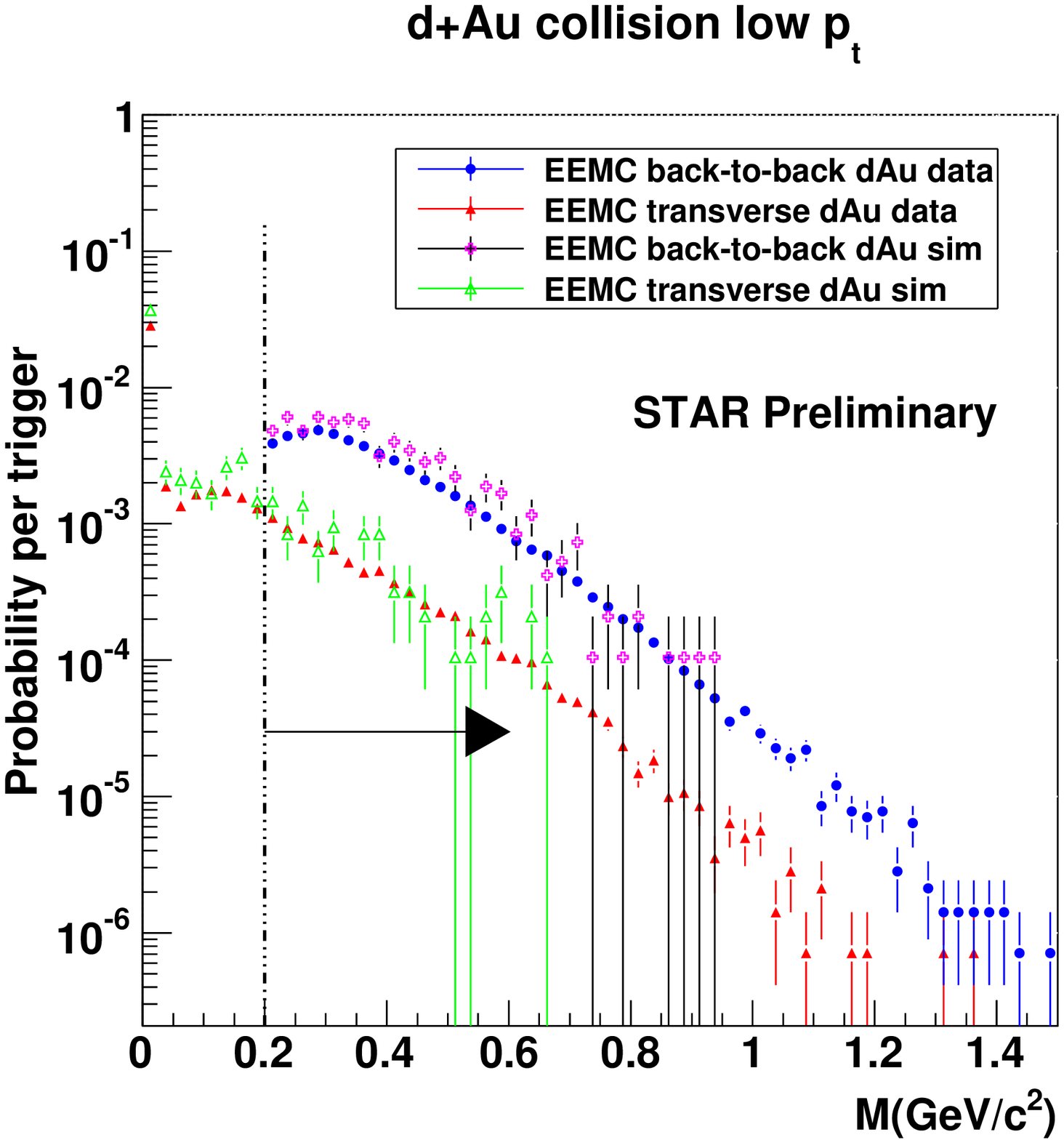}
\end{center}
\caption{Invariant masses of the EEMC jet-like cluster with $M > 0.2~\textrm{GeV/c}^{2}$ and $1.0~\textrm{GeV/c}<p_{t}^{EEMC}<p_{t}^{FMS}$ in the back-to-back region of the leading FMS $\pi^{0}$ with $p_{t}^{FMS} > 2.0~\textrm{GeV/c}$ in data (blue solid circles) and simulation (magenta open cross points). Underlying event with no mass and $p_{t}$ cuts in the transverse region in data (red solid triangle points) and simulation (green open triangle points) of p+p (left) and d+Au (rigtht) collisions.}
\label{EEMC_jet_mass}
\end{figure}

The apparent direction of the physical jet may be shifted by the underlying event contributions. Such contributions can arise from initial and final state interactions. The method which was developed for Tevatron high $p_{t}$ mid-rapidity data \cite{rick_med} is extended to RHIC forward rapidity low $p_{t}$ data. In events containing a FMS $\pi^{0}$ with $p_{t}^{FMS}>2.0~\textrm{GeV/c}$ and an EEMC jet-like cluster with $M>0.2~\textrm{GeV/c}^{2}$ and $1.0~\textrm{GeV/c}<p_{t}^{EEMC}<p_{t}^{FMS}$; when the azimuthal angle difference of them is within [$\frac{5\pi}{6}$, $\frac{7\pi}{6}$], the transverse region relative to the FMS $\pi^{0}$ ($\Delta\varphi$ within [$\frac{\pi}{3}$, $\frac{2\pi}{3}$] and [$\frac{4\pi}{3}$, $\frac{5\pi}{3}$]) is defined as the area for underlying event contributions. The underlying event candidates are reconstructed with jet cone algorithm. Comparisons of the invariant mass of back-to-back EEMC jet-like clusters ($M > 0.2~\textrm{GeV/c}^{2}$ and $1.0~\textrm{GeV/c}<p_{t}^{EEMC}<p_{t}^{FMS}$) and the invariant masses of underlying events (no $M$ and $p_{t}$ cuts) in data  and simulation of p+p and d+Au collisions are shown in Figure \ref{EEMC_jet_mass}. For mass above 0.2 $\textrm{GeV/c}^{2}$, the ratio of counts in the transverse region to the jet-like clusters is defined as a measure of underlying event contribution. The underlying event contribution decreases as the tower threshold (or mass lower limit for jet-like cluster) is increased. To suppress the underlying event contribution to the jet-like clusters, the EEMC tower threshold is chosen as 600 MeV and require the reconstructed jet-like cluster mass $>0.4~\textrm{GeV/c}^{2}$. 

\section{Correlation analysis and Systematics}
\begin{figure}[ht]
\begin{center} 
\includegraphics[width=0.96\textwidth]{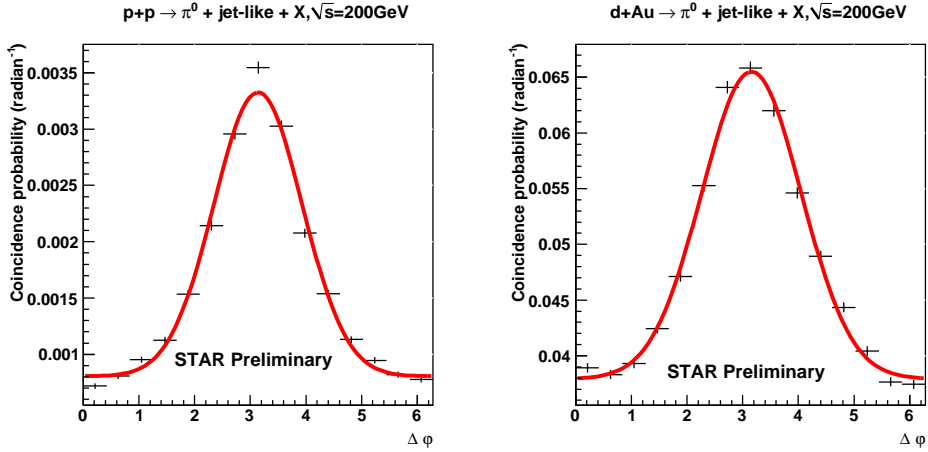}
\caption{The coincidence probability of FMS $\pi^{0}$ + EEMC jet-like cluster, as a function of relative azimuthal angle, with FMS $\pi^{0}$ $p_{t}^{FMS} > 2.0~\textrm{GeV/c}$ and EEMC jet-like cluster (mass $> 0.4~\textrm{GeV/c}^{2}$) $1.0~\textrm{GeV/c} < p_{t}^{EEMC} < p_{t}^{FMS}$ in p+p collisions and d+Au collisions. Data points are fitted with function $G(x)$ (see definition in Eq \ref{eq.corr_fit}) which is represented by the red solid line. Fitted width $\sigma_{pp} = 0.79\pm0.01$, and $\sigma_{dAu} = 0.89\pm0.02$. The errors are statistical only.}
\label{FMS_EEMC_final}
\end{center}
\end{figure}

The acceptance-corrected coincidence probability of FMS $\pi^{0}$ + EEMC jet-like cluster, as a function of relative azimuthal angle, with FMS $\pi^{0}$ $p_{t}^{FMS} > 2.0~\textrm{GeV/c}$ and EEMC jet-like cluster ($M > 0.4~\textrm{GeV/c}^{2}$) $1.0~\textrm{GeV/c} < p_{t}^{EEMC} < p_{t}^{FMS}$ is shown in Figure \ref{FMS_EEMC_final} for p+p and d+Au collisions. Data points are fit with function $G(x)$,
\begin{equation}\label{eq.corr_fit}
G(x) = A_{0} + \frac{A_{1}}{\sqrt{2\pi}\sigma} \times exp(-\frac{1}{2}(\frac{x-A_{2}}{\sigma})^{2}) ,
\end{equation}
where $A_{0}$ is the uncorrelated pedestal underneath the correlation peak, $A_{1}$ is the integral of the correlation peak, $\sigma$ is the width of the correlation and $A_{2}$ is the correlation peak centroid. Systematic studies show that the underlying event contribution can be suppressed by increasing tower threshold, while the correlation width differences between p+p and d+Au collisions do not change with different tower thresholds \cite{xuanli_thesis}. Figure \ref{FMS_EEMC_final}, where the tower threshold is set as 600 MeV, shows a width difference of $\sigma_{dAu}-\sigma_{pp} = 0.10 \pm 0.02$, the error is statistical only.

\begin{figure} 
\begin{center} 
\includegraphics[width=0.48\textwidth]{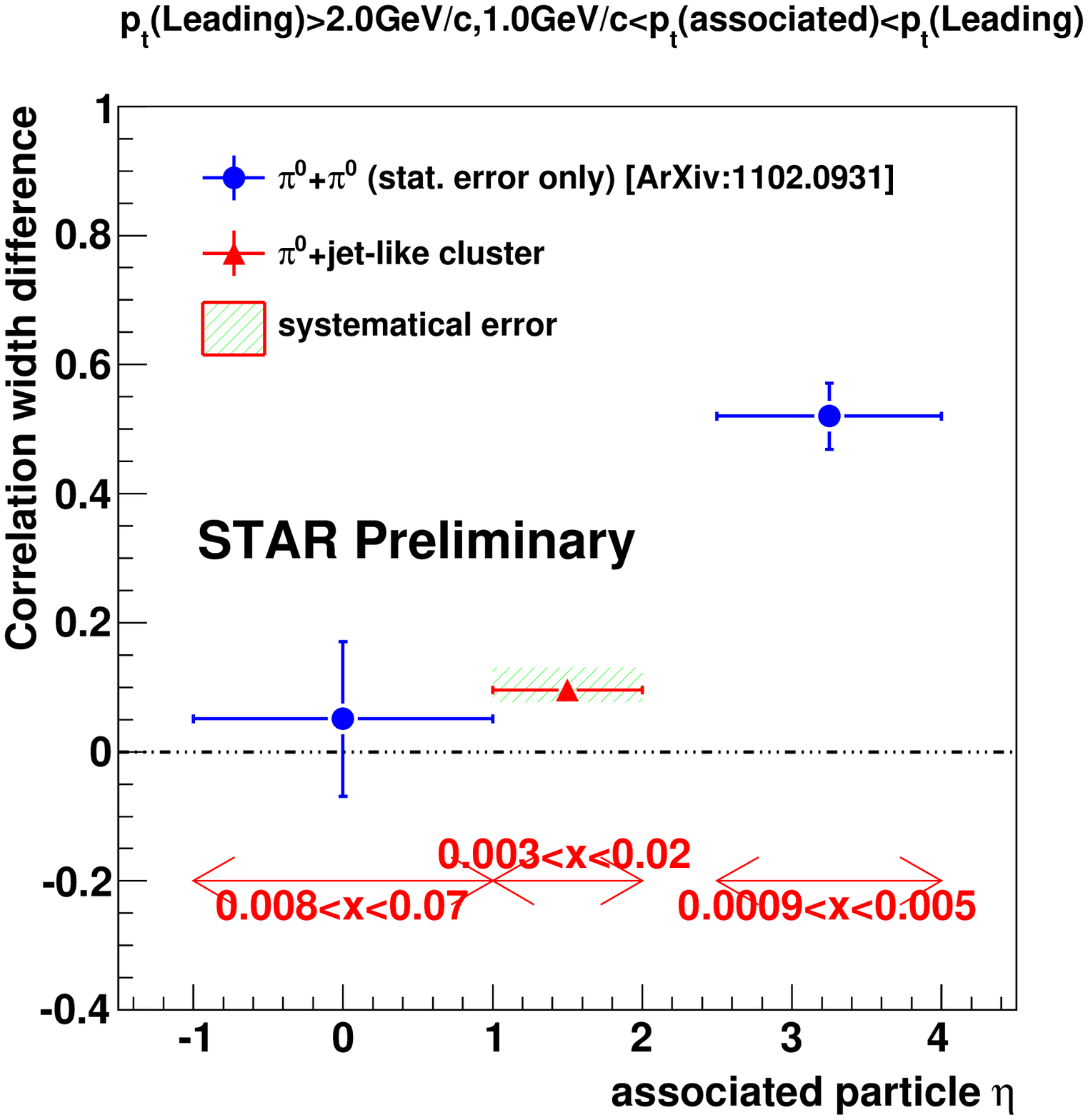}
\includegraphics[width=0.48\textwidth]{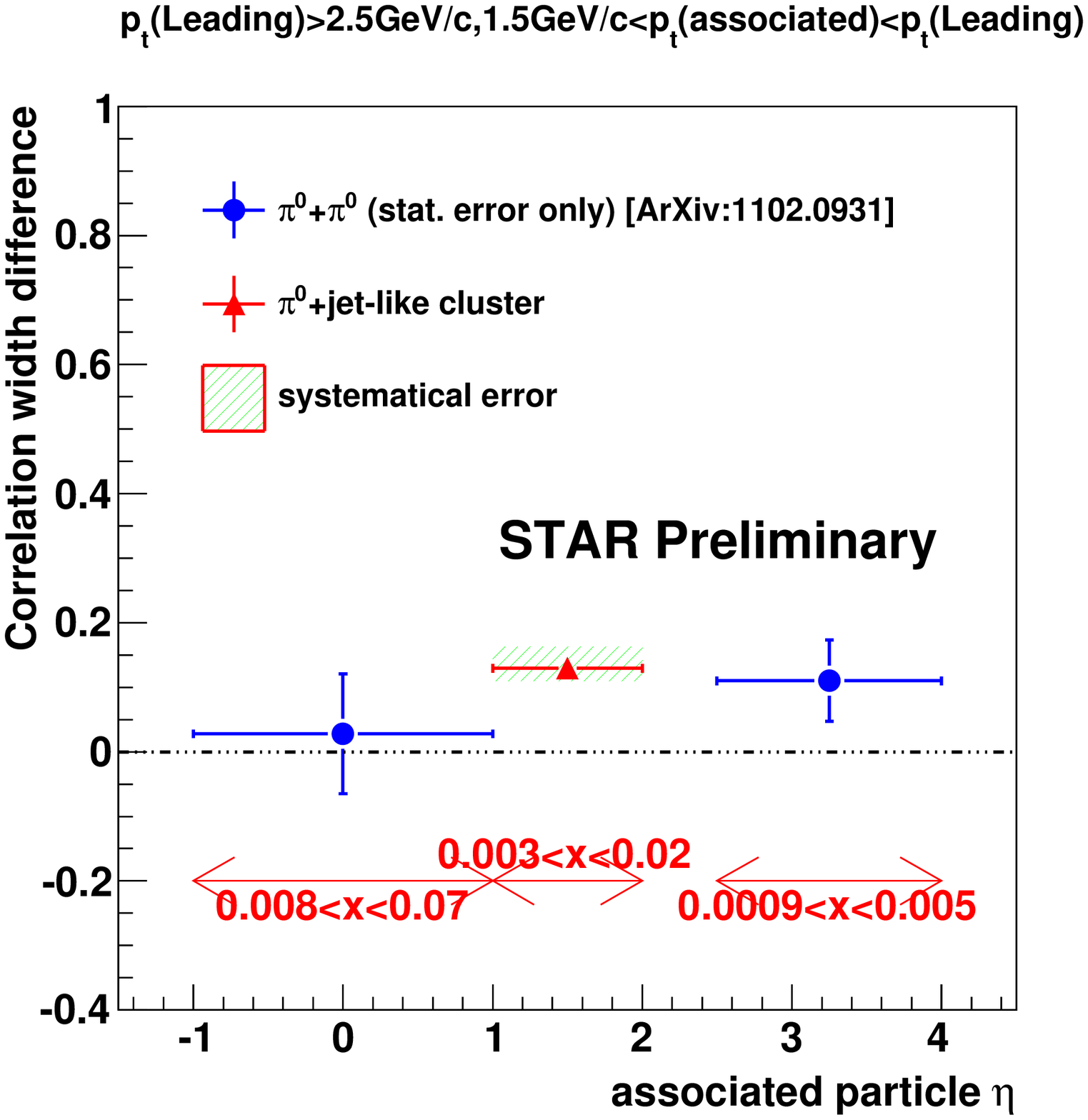}
\caption{Width differences between p+p and d+Au collisions versus the pseudorapidity of the associated particle in the forward di-hadron correlations at STAR. Left: Leading forward $\pi^0$ ($p_{t}^{Leading}>2.0~\textrm{GeV/c}$, $2.5<\eta^{Leading}<4.0$) and the associated $\pi^0$ and jet-like clusters with $1.0~\textrm{GeV/c}<p_{t}^{Associated}<p_{t}^{Leading}$. Right: Leading forward $\pi^0$ ($p_{t}^{Leading}>2.5~\textrm{GeV/c}$, $2.5<\eta^{Leading}<4.0$) and the associated $\pi^0$ and jet-like clusters with $1.5~\textrm{GeV/c}<p_{t}^{Associated}<p_{t}^{Leading}$. Blue circle points stand for $\pi^{0}$+$\pi^{0}$ azimuthal correlations with statistical errors, and red triangle points stand for $\pi^{0}$+jet-like cluster azimuthal correlations with statistical and systematic errors.}
\label{All_width_diff}
\end{center} 
\end{figure} 

Different selections of the cone radius, tower threshold, pseudorapidity and mass lower limit cuts for EEMC jet-like clusters can change the correlation results, and these effects are included in the systematic uncertainties. The final results on $\sigma_{dAu}-\sigma_{pp}$ of correlations between FMS $\pi^{0}$ with $p_{t}^{FMS}>2.0~\textrm{GeV/c}$ and EEMC jet-like cluster with $1.0~\textrm{GeV/c}<p_{t}^{EEMC}<p_{t}^{FMS}$ are $0.10 \pm 0.02^{+0.04}_{-0.02}$. For higher $p_{t}$ cuts (FMS $\pi^{0}$ with $p_{t}^{FMS}>2.5~\textrm{GeV/c}$ and EEMC jet-like cluster with $1.5~\textrm{GeV/c}<p_{t}^{EEMC}<p_{t}^{FMS}$), the width difference in the correlations is $\sigma_{dAu}-\sigma_{pp} = 0.13 \pm 0.02^{+0.03}_{-0.02}$. To summarize forward di-hadron correlation results at STAR, Figure \ref{All_width_diff} shows the width differences between p+p and d+Au collisions versus the pseudorapidity of the associated $\pi^{0}$ or jet-like cluster with different $p_{t}$ cuts \cite{ermes_thesis, for_for, xuanli_thesis}. Probed x regions for different measurements are shown in Figure \ref{All_width_diff} as well. As x decreases, the width differences between p+p and d+Au collisions increase.

\section{Summary}
The forward+near-forward correlations at STAR extends the probed x region to $0.003<x<0.02$. A smooth transition from a dilute parton gas to a saturated parton density is studied through the rapidity dependent forward di-hadron correlation studies at STAR. As the di-hadron correlations include both initial and final state interactions, a future Drell-Yan measurement may provide a cleaner channel to study the initial state.

\end{document}